\documentclass[aps,prl,twocolumn,showpacs,superscriptaddress,groupedaddress]{revtex4}  
\usepackage{graphicx}  
\usepackage{dcolumn}   
\usepackage{bm}        
\usepackage{amssymb}   

\usepackage[latin1]{inputenc}
\usepackage{tikz}
\usetikzlibrary{shapes,arrows}

\begin{document}

\widetext

\title{Event-Driven Monte Carlo: exact dynamics at all time-scales\\ for discrete-variable models}

\author{Alejandro Mendoza-Coto}
 \affiliation{Departamento de F\'\i sica, Universidade Federal do Rio Grande do Sul - CP 15051, 91501-970, Porto Alegre,
Brazil}

\author{Rogelio D\'iaz-M\'endez}%
\affiliation{icFRC, IPCMS (UMR 7504) and ISIS (UMR 7006), Universit\'e de Strasbourg and CNRS - 67000 Strasbourg, France}%

\author{Guido Pupillo}%
\affiliation{icFRC, IPCMS (UMR 7504) and ISIS (UMR 7006), Universit\'e de Strasbourg and CNRS - 67000 Strasbourg, France}%
\affiliation{FRIAS, Freiburg Institute for Advanced Studies - 79104 Freiburg, Germany}%

\date{\today}

\begin{abstract}
We present an algorithm for the simulation of the exact real-time dynamics of classical many-body
systems with discrete energy levels. 
In the same spirit of kinetic Monte Carlo methods, a stochastic solution of the master equation is found, with no
need to define any other phase-space construction.
However, unlike existing methods, the present algorithm does not assume any particular statistical distribution to perform
moves or to advance the time, and thus is a unique tool for the numerical exploration of fast and ultra-fast dynamical
regimes. 
By decomposing the problem in a set of two-level subsystems, we find a natural variable step size, that is well defined from
the normalization condition of the transition probabilities between the levels. 
We successfully test the algorithm with known exact solutions for non-equilibrium dynamics and equilibrium thermodynamical
properties of Ising-spin models in one and two dimensions, and compare to standard implementations of kinetic Monte Carlo
methods. 
The present algorithm is directly applicable to the study of the real time dynamics of a large class of classical markovian
chains, and particularly to short-time situations where the exact evolution is relevant.
\end{abstract}

\maketitle

\newcommand{\tbs}{\tau_{\mathrm{ED}}}
\newcommand{\po}{P_i^{o}(t)}
\newcommand{\pt}{P_i^{f}(t)}
\newcommand{\go}{\Gamma_i^{of}}
\newcommand{\gt}{\Gamma_i^{fo}}

%

\section{Introduction}

Many body classical models with discrete energy levels, such as Ising-spin systems, are particular examples of markovian
chains \cite{krapivsky10}, whose growing interest includes fields as diverse as condensed matter physics, biology
\cite{armond14} and economics \cite{jaeckel02}.
Despite intense research, exact results for these systems are rare in statistical physics, even for the most simple
Hamiltonians \cite{baxter08}. 
In this context, Monte Carlo (MC) numerical calculations are often considered as a fundamental benchmark for theories and
experiments~\cite{landau00}.

While MC simulations usually provide accurate results for static properties of interacting discrete-variable models, the
situation is different regarding their dynamical evolution, which, lacking a first-principles equation of motion as in 
continuous-variable systems, should be generally described by a stochastic master equation \cite{krapivsky10,glauber63}.
The latter expresses the probability distribution $\mathcal{P}(X,t)$ of a given state $X$ at time $t$,  in the
form\cite{kampen07}
\begin{equation}
\small
\frac{\partial \mathcal{P}(X,t)}{\partial t}=\sum_Y W(X|Y)\mathcal{P}(Y,t)-\sum_Y W(Y|X) \mathcal{P}(Y,t)
\label{mme}
\end{equation}
where $W(Y|X)$ is the transition rate from state $X$ to state $Y$, in units of inverse time.

In model with discrete variables, where the states form a numerable set, the common requirement for a dynamical MC algorithm
is to reach asymptotically the equilibrium state, where the master equation fulfills detailed balance \cite{landau00}.
As a result Monte Carlo algorithms are usually based on the equilibrium (e.g., time-independent) transition probabilities
between states, instead of the time-varying probabilities resulting from the general solution of equation~(\ref{mme}). 
The standard Monte Carlo step (MCS) that is used as the time step in most algorithms thus measures just the extent of random
exploration over the configuration space and has no direct relation with physical time. 
In general, this can result in significant deviations between the MC dynamics and the dynamical behavior described by the
master equation. 
However some equilibrium algorithms are known to reproduce successfully certain dynamical laws.
For example, this is the case of the Metropolis algorithm that predicts the $m\sim t^{1/2}$ scaling for the magnetization
{\it m} of the 2D Ising model after a subcritical quench~\cite{cugliandolo03}.  

So far, the most important bridge between MCS and physical time has been built by a class of algorithms usually
called dynamic or kinetic Monte Carlo (KMC)~\cite{landau00, kratzer09}.
KMC algorithms use the information about the transition rates $W(X|Y)$ to select the new updates, thus assigning to this
process a real time related to the inverse rates.
  
More specifically, KMC algorithms use the fact that the average time between two consecutive events in the system is of the
order of $\langle\Delta t\rangle\sim R(X)^{-1}$, where $R(X)=\sum_YW(Y|X)$ is the total sum of all rates of individual
processes the system can undergo from a given state $X$~\cite{kratzer09}.
Therefore, single time step is updated in a realistic way using a Poissonian distribution, by the
expression $\Delta t=-R(X)^{-1} \mathrm{log}(x)$, with $x$ being a uniformly distributed variable between $0$ and $1$.
This trick allows one to map the simulation steps with a real time that is physically meaningful, and has become the current
standard for numerical calculations of the dynamics of discrete-variable models.
The KMC step is then completed by the execution of the process that has been selected following a specific rule. 
The choice of this specific rule have produced different KMC schemes: the so called first-reaction method \cite{kratzer09},
for example, selects always the process with the fastest rate, while in the most commonly used BKL or
Gillespie algorithm~\cite{bkl75} the probability of selecting a process is a linear function of the rates.   

As can be inferred from the discussion above, all standard KMC methods follow a Markov chain kinetics, sampling correctly
from the (usually unknown) solution of the master equation, and so producing stochastic trajectories along the actual time
axis.
These single trajectories, however, are very accurate as far as time scales remain larger than $\langle\Delta
t\rangle$. 
At times of the order of consecutive events, 
trajectories are not expected to reproduce the exact solution of the master equation in the time axis.
This loss of accuracy at small times prevents, for example, the inclusion in the KMC dynamics of any time-dependent parameter
whose variation is of the order of $\langle\Delta t\rangle$. 
A reliable numerical technique capable of reproducing the master equation kinetics for fast and ultra-fast regimes is still
lacking.

In this work we present a new algorithm for addressing the latter problem, that is based on the numerical solution of the
master equation.
The main requirements are that
(i) the system can be decomposed into a set of $N$ two-level
subsystems, and (ii) any dynamical evolution is realized by sequential transitions within these individual subsystems. 
In the following we refer to these
$N$
transitions as {\it minimal processes}. 
Condition (i) is the standard form of any Hamiltonians with Ising-like spins,
however, 
it
can be also
made to apply to, 
e.g., classical mixtures on lattices or any generic Potts models. Condition
(ii) is equivalent to the well
known single-spin-flip update procedure, which is widely used for
dynamical calculations of discrete models.
As we show below, conditions (i) and (ii) can be 
fulfilled in
any model with discrete-variables.
We expect that the algorithm will be of particular value for, e.g., short-time critical dynamics of interacting classical
models \cite{albano11}, phase order kinetics \cite{bray02} and driven systems in oscillating fields \cite{park13}. 
However, its validity is not restricted
to physical systems, nor
to short times, 
and may in principle be used as an alternative to
Metropolis or
KMC simulations in a large number of markovian chains
of different nature.


\section{Event-Driven Algorithm}

Without loss of generality, in the following we present the algorithm in terms of Ising
spins.
The
idea behind
the scheme
is the following: given an initial configuration of the interacting spins, within the
characteristic time scale $\tbs$ (which is a priori unknown) associated with the flip of a single spin from that {\it
specific} configuration, the time evolution of the whole system is described by a set of $N$ independent reduced master
equations. 
By solving the latter, the exact time dependent probability  $P_i(\Delta t)$ for spin $i$ to flip is obtained
analytically for each spin $i$ at any time $\Delta t$, where $\Delta t$ is the time interval since the previous spin flip. In
turn, the condition $ \sum_iP_i(\tbs)=1$ defines the value
of $\tbs$ consistent with the single spin flip for that given configuration. Once $\tbs$ is defined, the algorithm proceeds
with evaluating all $P_i(\Delta t)$ at time $\Delta t=\tbs$ and uses them to update the configuration. This concludes a step
of the algorithm. The whole procedure is then repeated.

At the beginning of each step, we consider each spin $i$ occupying level $o$ while level $f$ is initially free, so that
$P^o_i(\Delta t=0)=1$ and $P^f_i(0)=0$, where $P_i^{o,f}(t)$ corresponds to the occupation probabilities of the two levels.
Within $\tbs$, 
these occupation probabilities 
fulfill the rate equations
\begin{eqnarray}\label{eq:rate}
\frac{d\po}{dt}&=&\gt\pt-\go\po \nonumber\\
\frac{d\pt}{dt}&=&\go\po-\gt\pt
\end{eqnarray}
where $\go$ and $\gt$ are the transition rates of the two-level subsystem of spin $i$, depending on the energy value
$E_i^{o,f}$ of the levels and the physical nature of the system. 
The limit of infinite time corresponds to Boltzmann occupation probabilities
$P^{o,f}_i(\infty)=e^{-\beta E_i^{o,f}}/Z$, where $Z$ is the partition function of the two-level
subsystem, $\beta^{-1}=k_BT$ and $k_B$ is the Boltzmann constant.

With these conditions the transition probability $P_i(\Delta t)$, i.e., from $o$ to $f$, can be written in the form
\begin{equation}
P_i(\Delta t)=P_i^f(\Delta t)=P_i^f(\infty)\left[1-e^{-(\go+\gt)\Delta t}\right],
\label{p2}
\end{equation}
where $P_i^f(\infty)=1/[1+e^{\beta(E_i^f-E_i^o)}]$.
As usually done in literature, in the following we assume that the characteristic frequency $\Gamma=\go+\gt$ is constant
in the system, and $1/\Gamma$ is adopted as the unit of time
\cite{glauber63}.
From Eq.~(\ref{p2}) and by applying the normalization condition $\sum_iP_i(\tbs)=1$ given above, we obtain $\tbs$ as
\begin{equation}
 \Gamma \tbs = - \mathrm{ln}\left[1-P^{-1}_*\right],
 \label{tbs}
\end{equation}
with $P_*=\sum_iP^f_i(\infty)$.

Each step of the  algorithm starts with the calculation, for each spin $i$, of the energy difference $\Delta
E_i=E_i^f-E_i^o$
associated to flipping the spin. From this, the value of $P_*$ can be calculated. If $P_*>1$ then the value of $\tbs$ for the
current step becomes that of expression Eq.~(\ref{tbs}).
If $P_*\leq1$, we choose to set $\tbs=1$ (see below).
Once $\tbs$ is evaluated, the sites are updated with the corresponding probability Eq.~(\ref{p2}) with $\Delta t=\tbs$,
resulting in an average of one spin flip. 
Consequently, the total time of the simulation is now incremented by $\tbs$.

A value of $P_*$ less than or equal to one, means that the system will never reach a time for which, in average, one
spin is flipped.
This is the well-known situation in which finite systems freeze, 
and the dynamics arrests, after reaching a stable configuration at sufficiently low temperatures [one example is shown below
when discussing Fig. \ref{f_eq} (inset)].
In most cases, this condition should suggest the end of the calculation, since the system will never evolve after reaching
this state. 
However, for problems in which the energy can change independently of the configuration (e.g., time-dependent Hamiltonians),
this freezing could be temporary and, consequently, $\tbs$ should be set to a constant value when $P_*\leq1$.
The value $\tbs=1$ is just a conventional number, since it has to be tailored to well-capture the time scale associated to
the energy changes in the problem at hand. 

We note that $\tbs$ corresponds to a discretization of  real time and in general varies from step to step, as it is
linked to the elementary changes of the system. 
In turn, the latter depends only on the microscopic interactions in the Hamiltonian and the specific spin configuration. 
Thus, since the whole algorithm directly deals with the exact real time, when conditions (i) and (ii) above are
satisfied we expect that the results of the numerical simulation will reproduce well those of the exact master equation at
all time scales. 

The role of $\tbs$ can also
be seen as a coarsening of the dynamics to the next physically meaningful time value,
calculated exactly, and not generated from a distribution function as in KMC schemes. 
This time coarsening represents the stochastic counterpart of that in event-driven molecular
dynamics approaches~\cite{poschel05}, thus corresponding to the waiting time connecting two consecutive (stochastic) events.
Therefore we refer to the above-described scheme as the Event-Driven algorithm (ED). 
This algorithm is composed of two serial loops of size $N$, firstly performing the calculation of $P_*$ and secondly
updating the minimal processes with the corresponding probability. 
Consequently the ED step is of complexity ${\cal O}(N)$, that is, it scales linearly with the number of two-level
subsystems,
which is the same as the Metropolis Monte Carlo step.

In general, for any discrete-variable markovian chain, starting from a given configuration, the dynamical step is a rule
selecting the next configuration among $N$ possible choices.
In terms of the ED scheme, the latter means that for any markovian chain one can build the set of $N$ two-levels equations.
The only input of the algorithm is the list of the $N$ energy differences corresponding to each one of the possible
choices of configurations.  
The main idea of the ED scheme relies on the very commonly used approximation that many coupled equations can be
decoupled for the very short time scale in which the system performs what we call a minimal process. 
Using this fact the algorithm finds the characteristic time $\tau_{ED}$ for which only one minimal process is likely to
happen. 

Consider, for instance, the $q$-levels Potts model, in which each spin can be in one of the $q>2$ available states.
For a system of $N_s$ spins, this model will imply a number of subsystems of $N=(q-1)N_s$, since, for any given
configuration, a minimal process consists in the transition of one of the $N_s$ spins to one of its $(q-1)$
available states.
Each of these $N$ possible transitions is identified with a minimal process by the ED algorithm, though for this model those
minimal processes corresponding to the same spin are excluded.
Thus, we just need to evaluate the energy difference associated to each of these $N$ transitions. 

In general, the number of minimal processes is not even forced to be constant along consecutive steps, as is the case for
example in the lattice gas model. 
The latter consists of particles that occupy certain positions in a lattice, and are able to move only to first-neighbouring
empty sites. 
The minimal processes here should be taken as the set of all single possible moves that particles can perform. 
For a very diluted configuration, this number of subsystems is then $N=ZN_p$, where $N_p$ and $Z$ are the number of particles
and the coordination of the lattice, respectively. 
However, when two particles become nearest neighbours, the number of minimal processes $N$ is reduced.

\begin{figure}[t]
\includegraphics[height=\columnwidth, angle=-90]{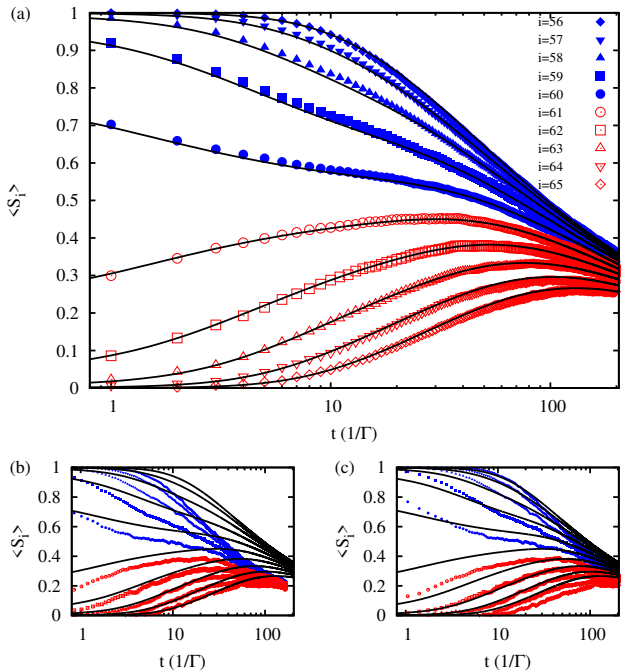}
\caption{Temporal evolution of the average local magnetization $\langle S_i(t)\rangle$ for ten different sites (from top to
bottom $i=56,57,58,\ldots,65$) of an Ising chain of $L=110$ sites at $T=0.1$.
The initial condition was set to $\langle S_i(0)\rangle=1$ for the ten central spins $51\leq i\leq60$, 
and
$\langle S_i(0)\rangle=0$ otherwise.
Dots are the outcome of numerical simulations using Event-Driven algorithm (a), and KMC algorithm fitting the data for large
(b) and small (c) spin index. 
Solid lines are the corresponding exact analytical solution obtained by R. J. Glauber in Ref. \cite{glauber63}. 
}
\label{f_glauber}
\end{figure}

\section{Numerical tests}

In the following we implement the ED algorithm.
Its accuracy is tested in a dynamical problem whose exact solution is known, and further compared to that of a
state-of-the-art $N$-fold KMC algorithm, implemented via the KMCLib library~\cite{kmclib14}.
Further tests are also presented to show the consistency of the ED scheme
with well known equilibrium and dynamical behaviors of the 2D Ising model, while discussing some specific features of the
method.

\subsection{Glauber exact solution}

We start by testing the algorithm in the exploration of the temporal evolution of the {\it local} magnetization in a linear
spin chain (1D) following a quench.
Before general tests involving averages for the total magnetization,  we compare here the predictions of the algorithm to the
exact solution of the full master equation, as obtained by Glauber~\cite{glauber63}. 
Up to our knowledge, this  remains the more complex discrete-variable statistical system for which the local magnetization
dynamics has been analytically obtained, in the full range of time scales and for arbitrary initial conditions. 
In turn, this analysis for the non-equilibrium properties of the local order parameter is the most complex test to which the
algorithm can be subjected.

Figure~\ref{f_glauber}a shows the time evolution of an Ising chain with $L=110$ sites, where the
initial state comprises a block of 10 parallel spins in the center, while the remaining 100 are in a disordered state
(see the caption for details).
The figure shows a perfect agreement between the exact analytical solution (continuous lines) and the numerical results from
ED algorithm.
Worth noting, this agreement occurs not only for the asymptotic, long-time regime, but also for very short times, where the
system is strongly out of equilibrium and the functional dependence of the local magnetization on time is non-trivial.
This confirms that the ED algorithm successfully accounts for the actual master-equation solution, accurately reproducing the
trajectories in the real time axis, even for scales of the order of single flips. 

For comparison, figures \ref{f_glauber}b and \ref{f_glauber}c shows the best fits for the outcome of the KMC algorithm
in the same problem.
By adjusting the time scale with a free parameter, a reasonable fit can be found at short times
for the local magnetization of sites far from (panel b), or deep into (panel c) the central ordered block of the initial
chain configuration.
While this rescaling is valid, it is impossible to find a single rescaling parameter successfully fitting all
the sites at once.
Moreover, as can be easily noticed from the figure, numerical and exact curves corresponding to sites near the edge of the
block, are completely impossible to collapse by solely a rescaling of the time axis.  

\subsection{2D Isind model}

We now focus on equilibrium properties. 
One important point is that, e.g. unlike KMC, here detailed balance is not directly used to determine the
transition probabilities, and in fact is in general not fulfilled. However, detailed balance is naturally recovered at
equilibrium in calculations.
\begin{figure}[t]
\includegraphics[height=\columnwidth,angle=-90]{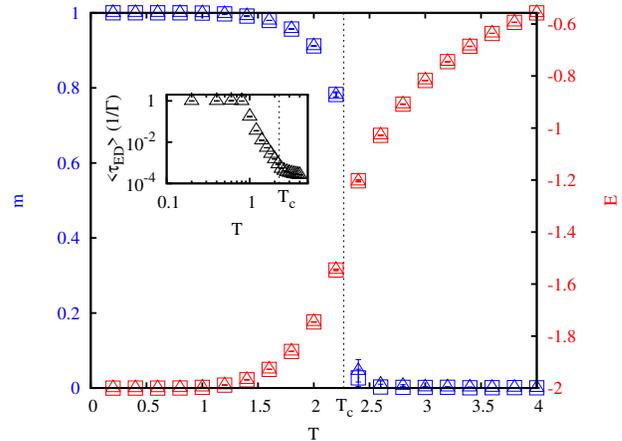}
\caption{Equilibrium magnetization $m$ (blue) and energy $E$ (red) obtained with the algorithm (triangles) and with
Metropolis calculations (squares), by a slow annealing of the system at different temperatures.
Units of energy and temperature are $J$ and $J/k_B$, respectively, where $J$ is the coupling constant of the Ising
Hamiltonian. 
Magnetization is the average value of the spins.
The dashed line corresponds to the exact transition temperature of the infinite system.
The inset is the temperature dependence of the average time step of the Event-Driven algorithm.}
\label{f_eq}
\end{figure}
Figure \ref{f_eq} shows example results for the equilibrium properties of the 2D Ising model with size $L^2=100\times100$
using both ED and Metropolis, equilibrated for $5\times10^4\Gamma^{-1}$ and $5\times10^4$MCS, respectively.
The system undergoes a phase transition from paramagnetic to ferromagnetic phase at $T_c=2.269$.  
The figure shows that the algorithm reproduces well the results from Metropolis for the magnetization and the  energy as a
function of $T$, finding the same equilibrium configurations and $T_c$.

While central to the algorithm, $\tbs$ can also capture certain interesting aspects of the system dynamics. 
In the inset of Fig.~\ref{f_eq}, the characteristic time $\tbs$, averaged over a time at least equal to the equilibration
one, is plotted as a function of $T$.  
For $T>T_c$, $\langle\tbs\rangle$ is very small ($\langle\tbs\rangle\sim 10^{-4}/\Gamma$), corresponding to a fast flipping
rate, as expected in the paramagnetic phase.
Below $T_c$, however, $\langle\tbs\rangle$ rapidly increases until it saturates for $T\lesssim1$. 
At this temperature, the dynamics is essentially frozen and $\tbs$ becomes one by construction. 
As discussed above, frozen dynamics is always reached in
calculations for
finite
systems evolving into a stable configuration (e.g., the ferromagnetic state).
This can often result in an unwanted slowing down of computations at sufficiently low $T$. A rapid growth of $\tbs$ (e.g.,
below $T_c$ in the figure) is then a computationally helpful flag of reaching a stable spin configuration.  
In fact, this is a limiting case of the time coarsening that is performed by $\tbs$ at each step of the algorithm, since
$\tbs$ is chosen to prevent spurious updating for $\Delta t<\tbs$ at each step.

We test the dynamical behavior of the algorithm for the 2D Ising model by quenching $T$ from a disordered configuration
(i.e., $T=\infty$) to a subcritical temperature $T=1<T_c$ corresponding to the fully magnetized ground state $m\approx1$ (see
Fig.~\ref{f_eq}). This is a well known coarsening process, where, as a result of quenching to low $T$, a mosaic of competing
ordered-phase clusters is formed.
In a finite system, the final state corresponds either to the fully ordered ground state or 
to a configuration with striped domains 
oriented antiparallel to the rest of the system \cite{olejarz12}. 
The two physically relevant times in this situation are the time 
$\tau_l$ 
associated with the appearence of the
first percolating cluster, i.e. an ordered domain of the size of the system, and the equilibration time $\tau_\mathrm{eq}$
after
which the system is found in one of the two final states.

\begin{figure}[t]
\includegraphics[height=\columnwidth,angle=-90]{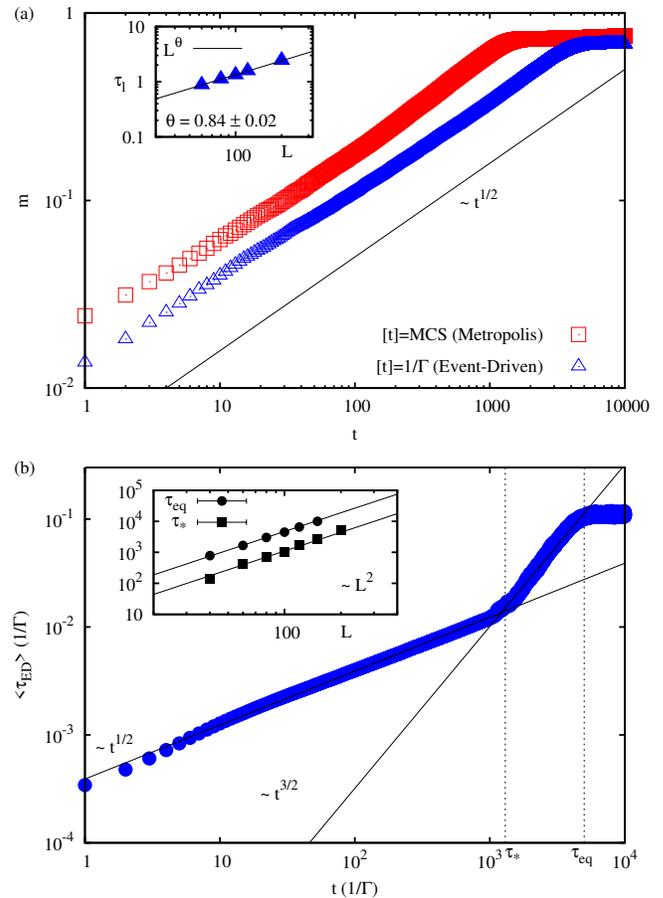}
\caption{(a) Evolution of the magnetization $m$ using ED (triangles) and Metropolis (squares), after a quench from
a disordered configuration into $T=1$,
for a system of $L^2=250\times250$.
Inset: size scaling of the characteristic time 
$\tau_l$ 
at which the first percolation
cluster is formed (see the text).
(b) Evolution of the average time step 
of the ED algorithm after the quench described in panel (a). 
Inset: size scaling of characteristic times $\tau_*$ and $\tau_\mathrm{eq}$ represented in the main figure (see the text).
}
\label{f_dyn}
\end{figure}

The evolution of the magnetization $m$ after the quench is shown in Fig.~\ref{f_dyn}a, where results are averaged over $500$
quench realizations in systems of up to $L^2=250\times250$ spins.
The figure shows that the ED algorithm reproduces the scaling $m\sim t^{1/2}$ typical of the coarsening dynamics of
two-dimensional systems with non-conserved order parameters \cite{cugliandolo03}, which is also captured by the Metropolis
dynamics, reaching the equilibrium configuration (i.e., plateau in the figure) at
$\tau_\mathrm{eq}\approx5\times10^3/\Gamma$.
The time 
$\tau_l$,
shown in the inset as a function of the 
linear size $L$,
signals the formation of the first percolating cluster, which has been demonstrated to be in general
unstable~\cite{cug14epl}.
Its computation was performed by first determining the ferromagnetic clusters, using an implementation of the Hoshen-Kopelman
algorithm~\cite{hk76}, 
and then by checking the percolating properties along the ED dynamics. 
We find an exponent $\theta=0.84$  for the power law 
$\tau_l\sim L^\theta$,
enriching the discussion on the  phase
order kinetics of models with non-conserved order parameter, usually developed within the KMC scheme~\cite{cug14epl}.

Further information on the quench dynamics is obtained by the time evolution of $\langle\tbs\rangle$  shown in
Fig.~\ref{f_dyn}b. 
Firstly, the equilibration time $\tau_\mathrm{eq}$ extracted from Fig.~\ref{f_dyn}a is well captured by the dynamics
of $\langle\tbs\rangle$. 
Consistently, the value $\langle\tbs\rangle\approx0.1/\Gamma$ for the plateau in Fig.~\ref{f_dyn}b is the same as that 
obtained at equilibrium for the corresponding temperature $T=1$ (see inset of Fig.~\ref{f_eq}).
In addition, (ii) new information  is provided by $\langle\tbs\rangle$  on the physical mechanisms of phase ordering. That
is, a second characteristic time-scale $\tau_*$ appears at $\tau_*\simeq 2 \times 10^3/\Gamma$, 
just where $\langle\tbs\rangle$ changes the slope. 
By inspection, we find that $\tau_*$ corresponds to the appearance of the first few stationary (i.e., final) states in some
realizations of the quenches.
That is, no final configuration is reached in our simulations for $t<\tau_*$. 
After this time, however, the system starts having a non-zero probability of being in the final state, where $<\tau_{ED}>$ is
maximal.
Consequently,  the average time scale of the relaxation slows down, in turn causing a more pronounced
slope.
In contrast, for $t>\tau_\mathrm{eq}$ all
configurations are either fully magnetized or striped. The inset shows that $\tau_\mathrm{eq}$ (as well as $\tau_*$) scales
with the system size as $\tau_\mathrm{eq}\sim L^2$, which is in agreement with known results \cite{cug14epl}.


\section{Conclusions}

In summary, we have introduced and tested a novel algorithm to simulate the stochastic dynamics of discrete variable models.
To the best of our knowledge, this is the first Monte Carlo method involving the exact physical time at all scales, with no
heuristic or phase-space assumptions.
The latter opens up the study of, e.g., strongly out-of-equilibrium situations for which exact numerical calculations are
currently not possible in short-time regimes. 

As said above, the present algorithm can be adapted to tackle several classes of different problems. 
For example, a microscopic update can be generalized that is consistent with conserved order parameter dynamics. The latter
can describe, e.g., the dynamics of kinetic phase separation in binary mixtures \cite{krapivsky10}.  
The role of two level subsystems is here played by each couple of nearest-neighbor sites with different occupations, while
minimal processes translate into exchanges within these subsystems. 
The same reasoning applies to general Potts models and related markovian chains. 
The study of quenches in classical many-body systems and the relation to Kibble-Zurek mechanism \cite{liu14}, Lieb-Robinson
bounds with short- and long-range interactions \cite{metivier14}, as well as dynamical phase transitions in magnetic models
\cite{berger13}, are other important examples of physical processes of current interest where our algorithm can be
straightforwardly applied.

\section{Acknowledgments}

We thank L. Nicolao and N. Prokofiev for useful discussions. 
We also acknowledge partial financial support from CNPq (Brazil), as well as the European Commission via ERC-St Grant ColdSIM
(No. 307688), EOARD, FWF-ANR grant "BLUESHIELD", EU via "RYSQ" and "COHERENCE", and UdS via Labex NIE and IdEX, Initial
Training Network COHERENCE and computing time at the HPC-UdS.

\bibliography{belen}

\end{document}